\documentclass[journal]{IEEEtran}

\ifCLASSINFOpdf
\else
   \usepackage[dvips]{graphicx}
\fi
\usepackage{url}

\usepackage{graphicx}
\usepackage{xcolor}
\usepackage{amsmath,amssymb}
\usepackage{hyperref}
\usepackage{algorithmic}
\usepackage[linesnumbered, ruled]{algorithm2e}
\usepackage{setspace}

\begin{document}

\title{A Deep-Bayesian Framework for Adaptive Speech Duration Modification}

\author{Ravi Shankar and Archana Venkataraman, \IEEEmembership{Member, IEEE}
\thanks{This work was supported by NSF CAREER award 1845430 (PI Venkataraman). The authors are with the department of Electrical and Computer Engineering at the Johns Hopkins University, Baltimore, MD 21218 USA (e-mail: rshanka3@jhu.edu, archana.venkataraman@jhu.edu).}}

\markboth{Journal of \LaTeX\ Class Files, Vol. 19, No. 8, August 2020}
{Shell \MakeLowercase{\textit{et al.}}: Bare Demo of IEEEtran.cls for IEEE Journals}
\maketitle

\begin{abstract}
We propose the first method to adaptively modify the duration of a given speech signal. Our approach uses a Bayesian framework to define a latent attention map that links frames of the input and target utterances. We train a masked convolutional encoder-decoder network to produce this attention map via a stochastic version of the mean absolute error loss function; our model also predicts the length of the target speech signal using the encoder embeddings. The predicted length determines the number of steps for the decoder operation. During inference, we  generate the attention map as a proxy for the similarity matrix between the given input speech and an unknown target speech signal. Using this similarity matrix, we compute a warping path of alignment between the two signals. Our experiments demonstrate that this adaptive framework produces similar results to dynamic time warping, which relies on a known target signal, on both voice conversion and emotion conversion tasks. We also show that our technique results in a high quality of generated speech that is on par with state-of-the-art vocoders. 
\end{abstract}


\IEEEpeerreviewmaketitle

\section{Introduction}

\IEEEPARstart{H}{uman} speech is a rich and varied mode of communication that encompasses both language/semantic information and the mood/intent of the speaker. The latter is primarily conveyed by prosodic features, such as pitch, energy, and speaking rate. There are many applications where understanding and manipulating these prosodic features is required. Consider voice conversion systems. Pitch and energy modifications are used to inject emotional cues into the speech or to change the overall speaking style~\cite{face_vocal_expression, psych, diffeomorphic_hnet, hnet_max_likelihood, mellotron}. Prosodic features are also used to evaluate the quality of human machine dialog systems~\cite{prosody_eval}, and they play a significant role in speaker identification and recognition systems~\cite{voice_quality}. 

While there are many approaches for automated pitch and energy modification~\cite{global_variance_GMM, gmm_emo_conv, cyclegan_vc, vcgan, chained_model}, comparatively little progress has been made in changing the speaking rate of an utterance. In fact, the speaking rate plays a crucial role in conveying emotion~\cite{emotion_perception_factors} and in diagnosing human speech pathologies~\cite{stutter_evaluation}. The speaking rate is difficult to manipulate because, unlike pitch or energy, there is no explicit coding for the signal duration. Rather, it is implicitly defined by a collection of frame-wise spectral representations (e.g., the short time Fourier transform or Mel-frequency cepstral coefficients). As a result, duration modification algorithms are not adaptive; they either require considerable user supervision, or they are geared towards aligning two known speech signals.

Perhaps the earliest duration modification method is the time-domain phase overlap add (TD-PSOLA) algorithm~\cite{tdpsola}. TD-PSOLA modifies the pitch and duration of a speech signal by replicating and interpolating between individual frames. However, the user must manually specify both the portion of speech to modify and the exact manner in which it should be altered. Hence, the method is neither automated nor adaptive. An alternative approach is dynamic time warping (DTW), which finds the optimal time alignment between two parallel speech utterances~\cite{dtw}. DTW constructs a pairwise similarity matrix between all frames of the two utterances and estimates a \textit{warping path} between the starting $(0,0)$ and ending $(T_s,T_t)$ points of the utterances based on a Viterbi-like decoding of the similarity matrix. While simple, DTW requires both the source and target utterances to be known \textit{a priori}. Hence, it cannot be used for on-the-fly modification of new signals.

Finally, recent advancements in deep learning have led to a new generation of neural vocoders, which disentangle the semantic content from the speaking style~\cite{wavenet, tacotron, latent_uncovering}. These vocoders can alter the speaking rate via the learned style embeddings. While these models represent seminal contributions to speech synthesis, the latent representations are learned in an unsupervised manner, which makes it difficult to control the output speaking voice. Another drawback of these methods is the computational overhead and data resources required to train the models and generate new speech~\cite{analysis_speech_synthesis}. 

In this paper, we introduce the first fully-automated adaptive speech duration modification scheme. Our approach combines the representation capabilities of deep neural networks with the structured simplicity of dynamic decoding. Namely, we model the alignment between a source and target utterance via a latent attention map; these maps are used as the similarity matrix for backtracking. We train a masked convolutional encoder-decoder network to estimate these attention maps using a stochastic mean absolute error (MAE) formulation. We demonstrate our framework on a voice conversion task using the CMU-Arctic dataset~\cite{cmu_arctic} and on three multi-speaker emotion conversion tasks using the VESUS dataset~\cite{vesus}. Our experiments confirm that the proposed model can perform open-loop duration modification and produces high-quality speech. Finally, our approach differs fundamentally from the conventional DTW~\cite{dtw} algorithm which requires both, the source and target utterances to warp one onto the other.

\vspace{-2mm}
\section{Method}
\begin{figure}[t]
  \centering
  \includegraphics[width=0.65\linewidth]{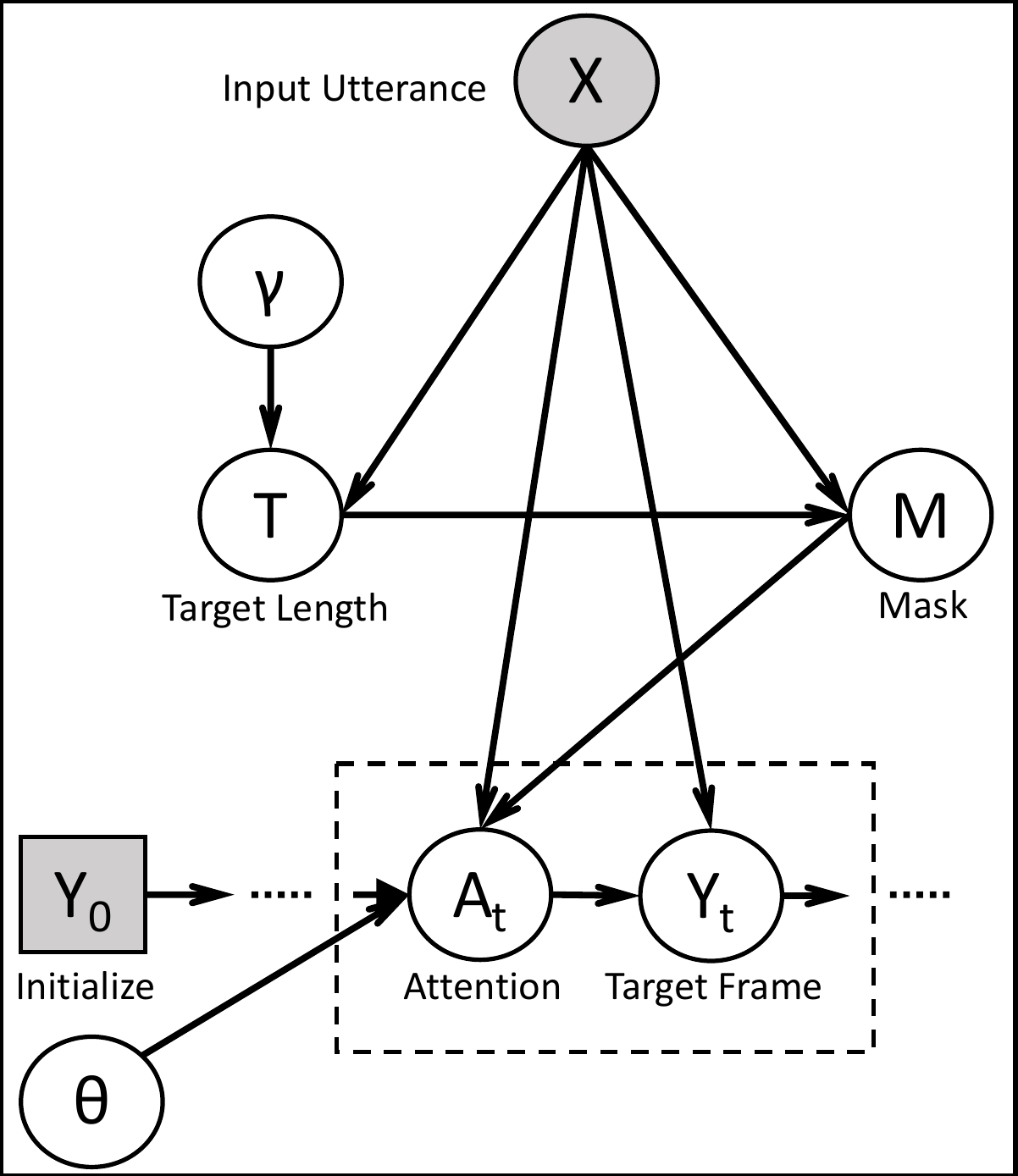}
  \caption{Graphical model for duration modification. $\gamma$ and $\theta$ are the model parameters which are inferred during training.}
  \label{fig:graphical_model}
  \vspace*{-2mm}
\end{figure}

Fig.\ref{fig:graphical_model} illustrates our underlying generative process. 
Given an utterance $X$, we first estimate the length $T$ of the (unknown) target utterance $Y$ and subsequently use it to estimate a mask $M$ for the attention map. The mask restricts the domain of the attention vectors $A_t$ at each frame~$t$ to mitigate distortion of the output speech. We use paired data $(X_{tr},Y_{tr})$ to train an encoder-decoder network to generate the attention vectors. During testing, we first generate the attention map from the input~$X$ and use it to produce the target speech~$Y$.

\vspace{-2mm}
\subsection{Loss Function}
Let $X \in \mathbb{R}^{D \times T_s}$ denote the input speech. In this work, $X$ corresponds to the filter-bank energies, where $D$ is the number of filter-banks, and $T_s$ is the number of temporal frames in the utterance. Similarly, we denote target speech as $Y \in \mathbb{R}^{D \times T}$. Notice that the target utterance length $T$ may differ from $T_s$.

Our generative process for the target speech is as follows:
\begin{equation}
T \sim \text{Laplace} (T^0, b_T) \;\;\; and \;\;\; Y_t \sim \text{Laplace} (Y_t^0, b_y),
\label{eqn:T_y_modeling}
\end{equation}
where $T$ is the length of the target utterance, and $Y_t$ is the target features at time~$t$. The parameters~$\{T^0, b_T, Y_t^0, b_y\}$ of the distributions are unknown and we implicitly estimate them via a deep neural network parameterized by $\gamma$ and $\theta$. 

By treating the unknown parameters as functions of the input~$X$, we obtain the following estimating equations for the target sequence length and frame-wise filter-bank energies:
\begin{equation}
\hat{T} = f_\gamma(X) \;\;\; and \;\;\; \hat{Y}_t = X \cdot A_t + f_\theta(X, \hat{Y}_{0:t-1}).
\label{eqn:T_y_estimation}
\end{equation}
The functions $f_\gamma(\cdot)$ and $f_\theta(\cdot,\cdot)$ correspond to deep networks. The variable $A_t \in \mathbb{R}^{T_s}$ is an attention vector that combines frame-wise features of the source utterance~$X$ to generate the target frame~$\hat{Y}_t$. Notice that the residual, which cannot be explained by the input utterance, depends on the predictions $\hat{Y}_{0:t-1}$ at previous time steps. This autoregressive property allows the neural network to learn a time-varying component that can differentiate between the speakers or emotions. 

During training, we use paired data $(X,Y)$ to maximize the likelihood of the target speech signal with respect to the neural network weights~$\{\theta,\gamma\}$. This likelihood can be written as:
\begin{equation}
P(\hat{Y}, \hat{T} | X) = P(\hat{T} | X) \prod_{t=1}^{\hat{T}} P(\hat{Y}_t | X, \hat{T}, \hat{Y}_{0:t-1}),
\label{eqn:joint_lkl}
\end{equation}
where, the second term of Eq.~(\ref{eqn:joint_lkl}) can be expanded as follows:
\begin{align} \nonumber
P(&\hat{Y}_t | X, \hat{T}, \hat{Y}_{0:t-1}) = \sum_{A_t} P(\hat{Y}_t, A_t | X, \hat{T}, \hat{Y}_{0:t-1}, M) \\ \label{eqn:mask_attention}
& = \sum_{A_t} P(\hat{Y}_t | X, \hat{T}, A_t, \hat{Y}_{0:t-1}) P(A_t | X, \hat{Y}_{0:t-1}, M) 
\end{align}
The variable $M$ in Eq.~(\ref{eqn:mask_attention}) denotes the attention mask and is introduced for convenience; it is a deterministic function of the source speech length~$T_s$ and the estimated target length~$\hat{T}$. 

\begin{figure*}[!t]
  \centering
  \includegraphics[width=0.85\textwidth, height=7.5cm]{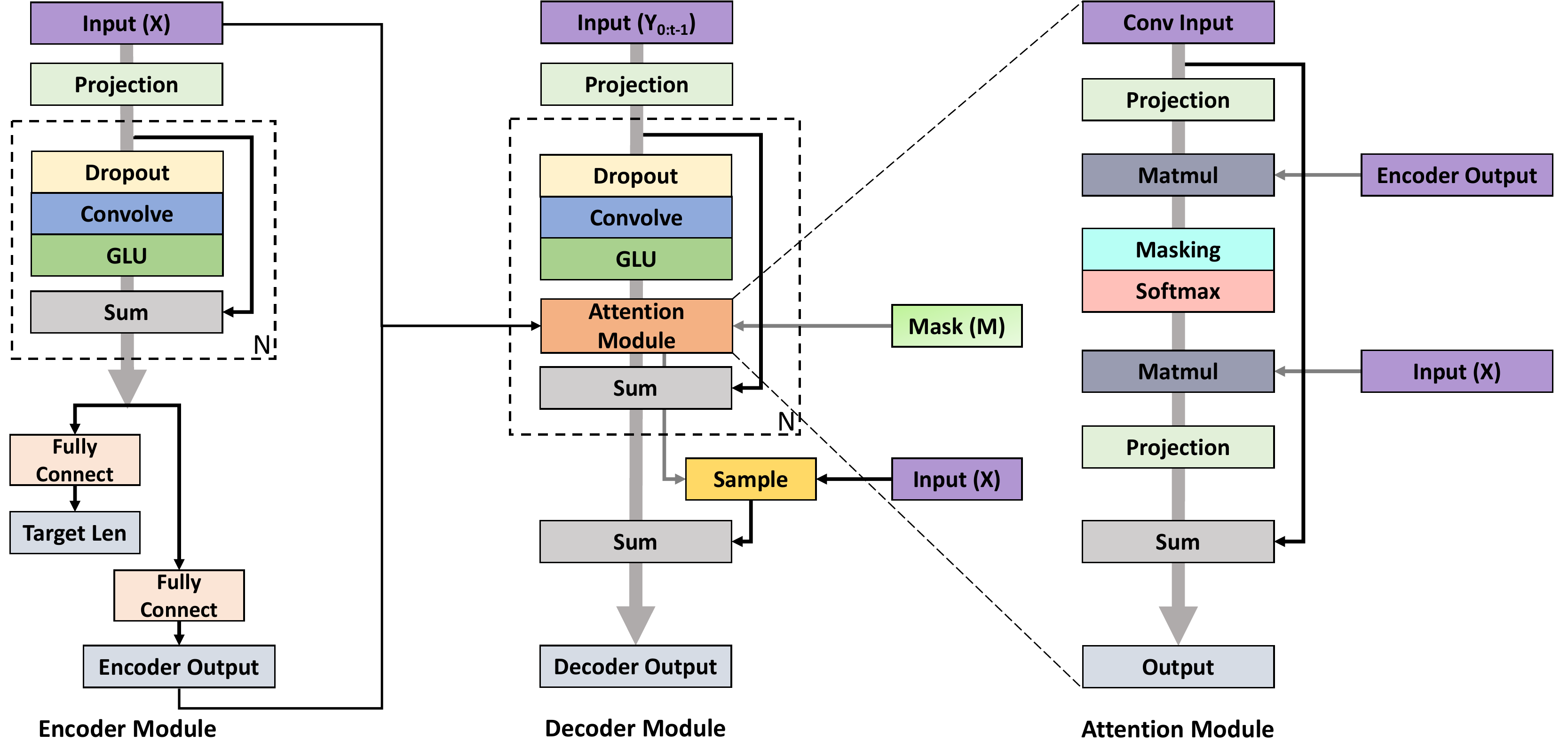}
  \caption{Neural network architecture used for the sequence-to-sequence speech generation. The encoder and decoder modules consist of 10 identical blocks. Projection layers are simple feed-forward layers without any non-linearity to project features in high dimension.}
  \label{fig:neural_network}
  \vspace*{-0.25cm}
\end{figure*}

We use a variational free energy formulation~\cite{variational_bayes} to derive an upper bound to our data log-likelihood (see supplemental materials for complete derivation). This bound can be translated into the following neural network loss function:
\begin{align} \nonumber
L &= E_{A_t \sim q_\theta} \big[ \log \big( P(\hat{Y}_t | X, A_t, \hat{Y}_{0:t-1}) \big) \big] + \log\big( P(\hat{T} | X) \big) \\ \label{eqn:expected_loss}
& = \lambda_1 \times E_{A_t}\big[ \Arrowvert \hat{Y}_t - Y_t^0 \Arrowvert_1 \big] \; + \;  \lambda_2 \times \Arrowvert \hat{T} - T^0 \Arrowvert_1
\end{align}
Here, $\lambda_1$ and $\lambda_2$ are model hyperparameters and implicitly contain the variances of Laplace distributions in Eq.~(\ref{eqn:T_y_modeling}). The distribution $q_\theta$ is a variational distribution which is approximated by the fully convolutional neural network in Fig.~\ref{fig:neural_network}.

\begin{figure}[t]
  \centering
  \includegraphics[width=0.9\linewidth, height=2.5cm]{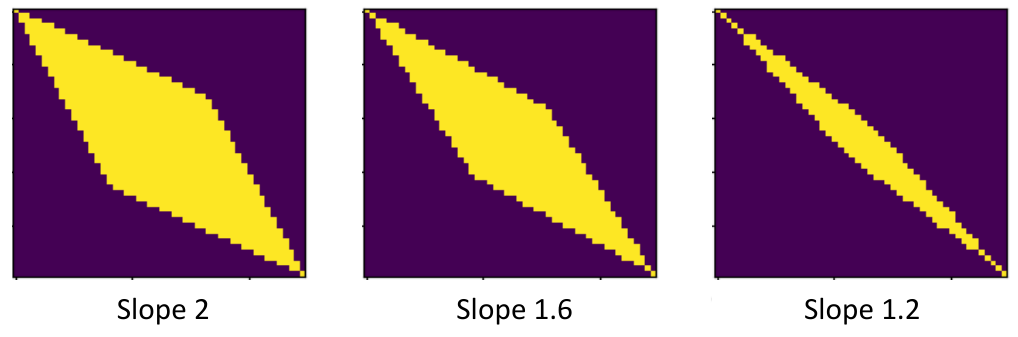}
  \caption{Binary attention masks with 3 different slopes.}
  \label{fig:masks}
  \vspace*{-4mm}
\end{figure}

\vspace{-3mm}
\subsection{Masking}
The mask~$M$ 
is used to constrain the scope of the attention mechanism to be similar in time-scale to the input. This procedure is important for two reasons. From a speech quality perspective, large swings in speaking rate may generate unintelligible speech. From an estimation perspective, the utterances contains hundreds (sometimes thousands) of frames. It is difficult to robustly train a deep network 
to generate such long attention vectors using smaller datasets.

We use the masks derived from Itakura parallelogram~\cite{itakura}, as illustrated in Fig.~\ref{fig:masks}. The Itakura parallelogram is used to speedup DTW algorithms when the speaking rates in the source and target utterances are expected to be similar~\cite{itakura}. The slope of the Itakura parallelogram specifies the minimum and maximum speaking rates that the reconstructed utterances are allowed to possess in comparison to the input speech. 

\begin{algorithm}[!t]
\setstretch{1.1}
    \SetKwInOut{Input}{Input}
    \SetKwInOut{Output}{Output}
    function \underline{modifyDuration} $(X)$\;
    \Input{filter-bank energy ($X \in \mathbb{R}^{D \times T_s}$ and $Y_0$)}
    \Output{alignments ($(x_1,y_1), (x_2,y_2), ... $)}
    Predict length of target sequence $T_t = f_\gamma(X)$\;
    Create attention mask $M \in \mathbb{R}^{T_s \times T_t}$ and Set $t = 0$\;
    \If{$ t < T_t $}
      {
        Using mask $M_t$, $X$, and $Y_{0:t-1}$ estimate $A_t$\;
        Using $X$, $Y_{0:t-1}$, and $A_t$, predict $Y_t$\;
        $t \leftarrow t+1$\;
      }
      {
        Run DTW backtracking on the attention matrix $A$\;
        return (alignments $(x_1,y_1), (x_2,y_2), ... (x_n, y_n)$)\;
      }
    \caption{Strategy for duration modification}
    \label{alg:duration_modification_strategy}
\end{algorithm}

\vspace{-3mm}
\subsection{Neural Network Architecture}
We adapt the neural network architecture from~\cite{conv_seq2seq} by adding skip connections to the last layer and changing the configuration of the attention module. Fig.~\ref{fig:neural_network} shows the encoder, decoder and the new attention module of the convolutional neural network. The encoder is responsible for generating feature embeddings for the decoder and for predicting the relative length of target speech. The sample operation in Fig.~\ref{fig:neural_network} is responsible for generating a sample from the attention distribution required for reconstruction and backpropagation. 

\begin{figure}[t]
  \centering
  \includegraphics[width=0.65\linewidth, height=3.8cm]{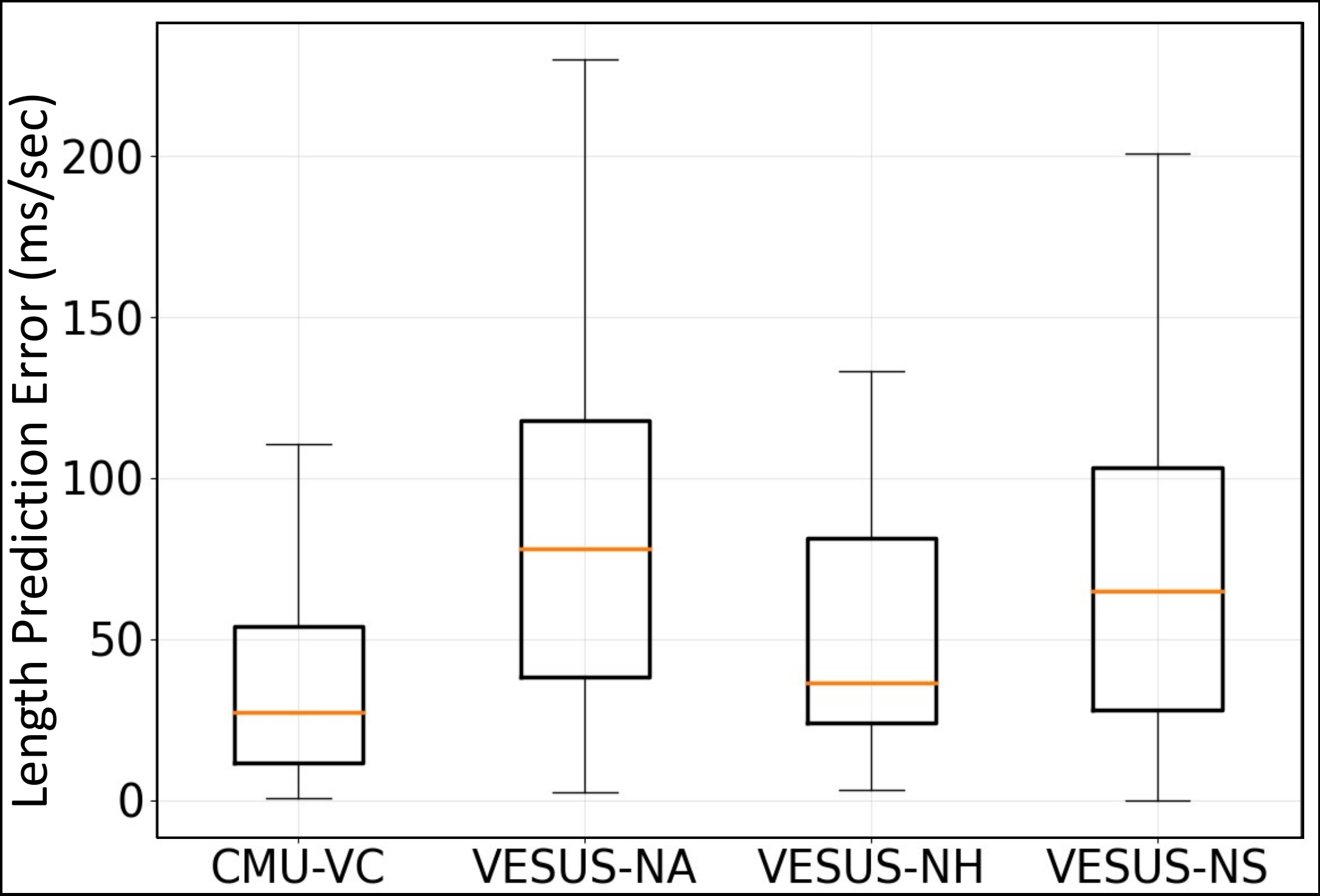}
  \caption{Error in length prediction using encoder embeddings.}
  \label{fig:len_pred}
  \vspace*{-0.30cm}
\end{figure}
\vspace{-0mm}

We train our model using the Adam optimizer~\cite{adam_optimizer} with a fixed learning rate of $10^{-4}$. The input $X$ is an 80-dimensional vector of Mel-filterbank energies. The projection layer expands this input to $256$ dimensions. Both the encoder and decoder consist of $10$ convolutional layers, each followed by gated linear unit. We use data augmentation to stabilize the network. Specifically, we reverse the input-output sequences and randomly extract intervals (with probability $0.5$) from the full utterance. Our full model training procedure is described in the supplementary materials. The source code can be download from:
\url{https://engineering.jhu.edu/nsa/links/}.

\vspace{-3mm}
\subsection{DTW Back-Tracking}
Our final step is to use the attention map produced by the decoder as a proxy for the DTW similarity matrix between the source and target speech frames. Effectively, we use the robust dynamic programming operation to get a path of alignment within the mask boundary, rather than rely on the noisy spectral reconstruction~(see Algorithm~\ref{alg:duration_modification_strategy}). To avoid skipping phonemes, the path is constrained to take at most one horizontal or vertical step consecutively while backtracking. We finally use this alignment as a lookup table to synthesize the target speech from the input via the WORLD vocoder~\cite{world_vocoder}. 

\vspace{-2mm}
\section{Experimental Results}
We evaluate our model on two multi-speaker datasets: CMU-ARCTIC~\cite{cmu_arctic} and VESUS~\cite{vesus}. We query three properties of our model on four tasks, as described below.


\vspace{-3mm}
\subsection{Data and Voice Morphing Tasks}
CMU-ARCTIC has 4 American English speakers (two male, two female), who are paired according to gender for voice conversion. We train our duration modification framework using $2164$ utterances from the database and use the remaining $100$ utterances for testing the open-loop modification properties. 

VESUS is an emotional speech corpus containing $250$ phrases read by $10$ speakers in $4$ emotion classes: neutral, angry, happy, and sad. VESUS also contains crowd-sourced emotional annotations. Here, we primarily use those utterances that are correctly annotated by at least half of the listeners.

We train three duration models corresponding to the three neutral-emotional pairs. This results in the following split:
\begin{itemize}
    \item \textbf{Neutral to Angry Conversion}: 2385 utterances for training, 72 for validation and, 61 for testing.
    \item \textbf{Neutral to Happy Conversion}: 2431 utterances for training, 43 for validation and, 43 for testing.
    \item \textbf{Neutral to Sad Conversion}: 2371 utterances for training, 75 for validation and, 63 for testing.
\end{itemize}
Given the small sample size due to shorter sequences, we fine-tune the model trained on CMU-ARCTIC for each emotion conversion task \textit{in lieu} of training the networks from scratch.

\begin{figure}[t]
  \centering
  \includegraphics[width=0.65\linewidth, height=3.8cm]{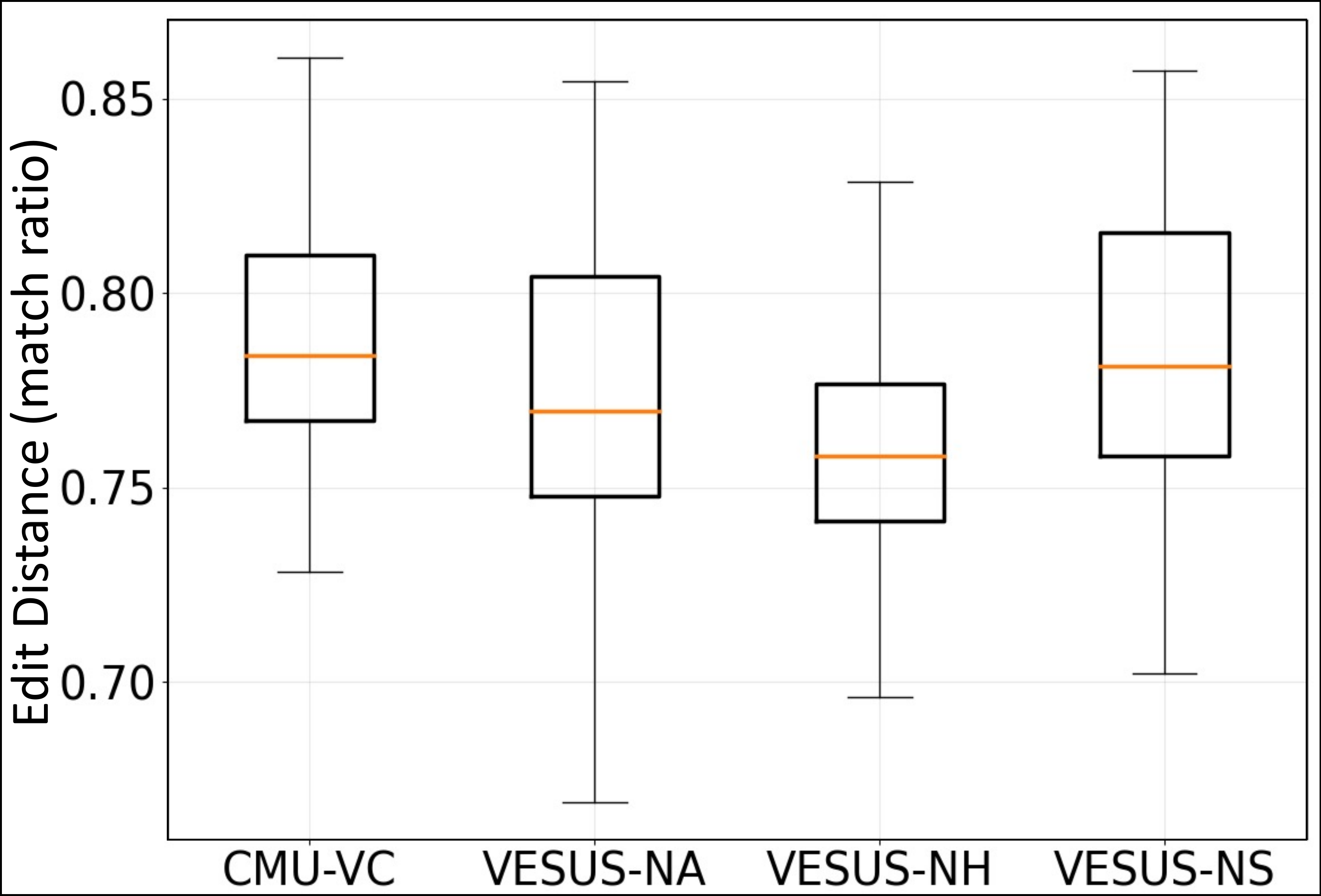}
  \caption{Alignment similarity between our method and DTW.}
  \label{fig:kl_div}
  \vspace{-2mm}
\end{figure}

\begin{figure}[t]
  \centering
  \includegraphics[width=0.65\linewidth, height=3.8cm]{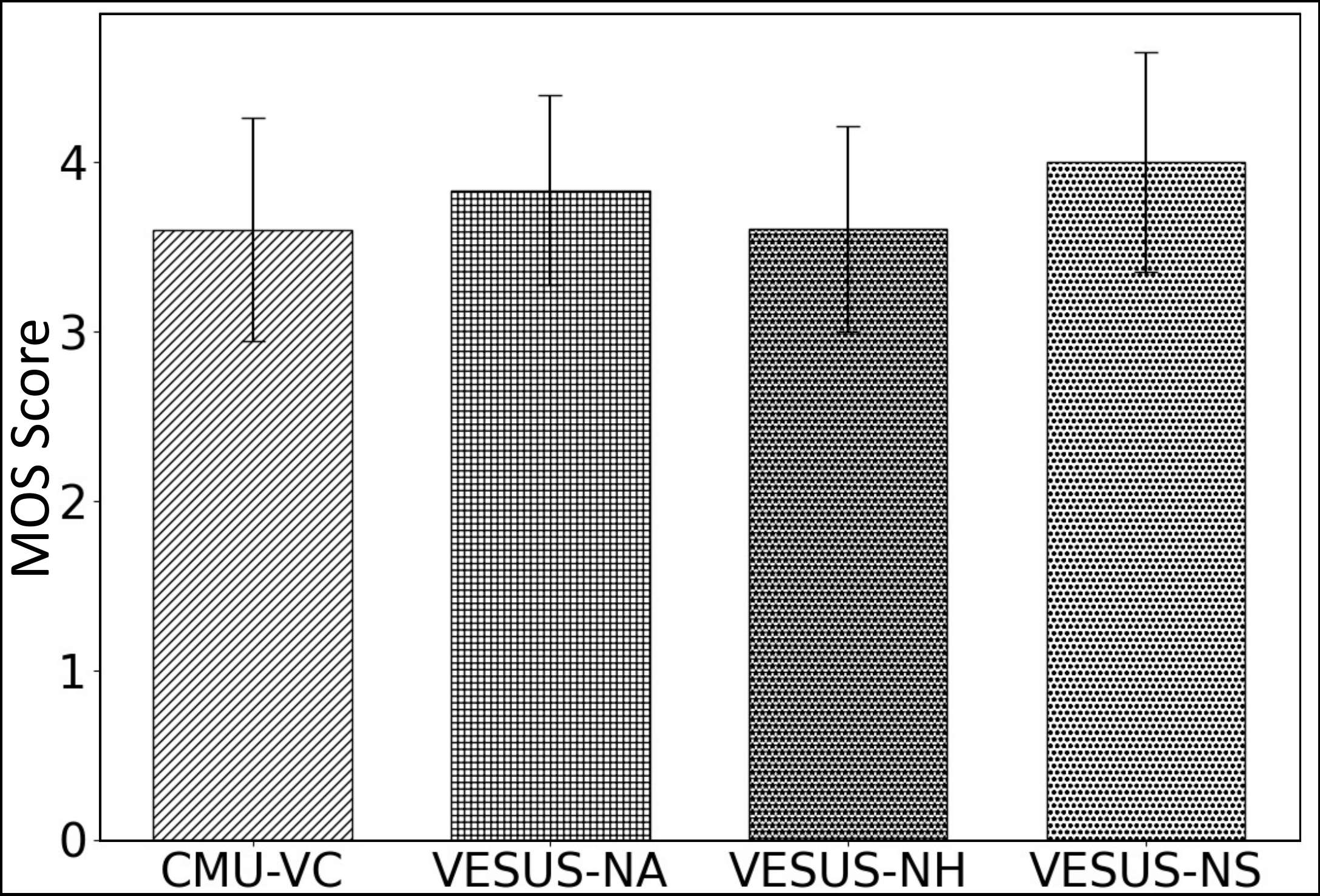}
  \caption{MOS of speech generated by our model evaluated via AMT.}
  \label{fig:mos_score}
  \vspace*{-2mm}
\end{figure}

\vspace{-3.5mm}
\subsection{Length Prediction}
As seen in Fig.~\ref{fig:neural_network}, we use the encoder embeddings to predict the length of the target utterance as a ratio of the source utterance length. Fig.~\ref{fig:len_pred} shows the error in predicting this ratio in a ms/sec. Notice that our framework mispredicts the utterance lengths by only $40$ms/sec and $65$ms/sec on CMU-ARCTIC and VESUS, respectively. Duration prediction is particularly challenging on VESUS due to marked differences between neutral and emotional utterances. However, our framework performs well even in this challenging scenario, likely due to our fusion of deep representation and Bayesian regularization.

\vspace{-3.5mm}
\subsection{Attention Alignment}
Next, we compare the alignment between source and target speech frames using our method with the original DTW algorithm. Recall that DTW requires access to the target speech utterance, whereas our approach does not. To compare the warping paths, we code the horizontal, diagonal, and vertical moves of the backtracking procedure into three classes. We then compute the edit distance between the DTW alignment and the attention map based alignment. Fig.~\ref{fig:kl_div} illustrates the match ratio normalized by the average length. As seen, the match ratio varies between $0.70$ and $0.85$, which suggests that our approach captures the general characteristics of an unseen target utterance. To our knowledge, this is the first demonstration of an adaptive duration modification framework. 

\begin{figure}[t]
  \centering
  \includegraphics[width=0.8\linewidth, height=5.5cm]{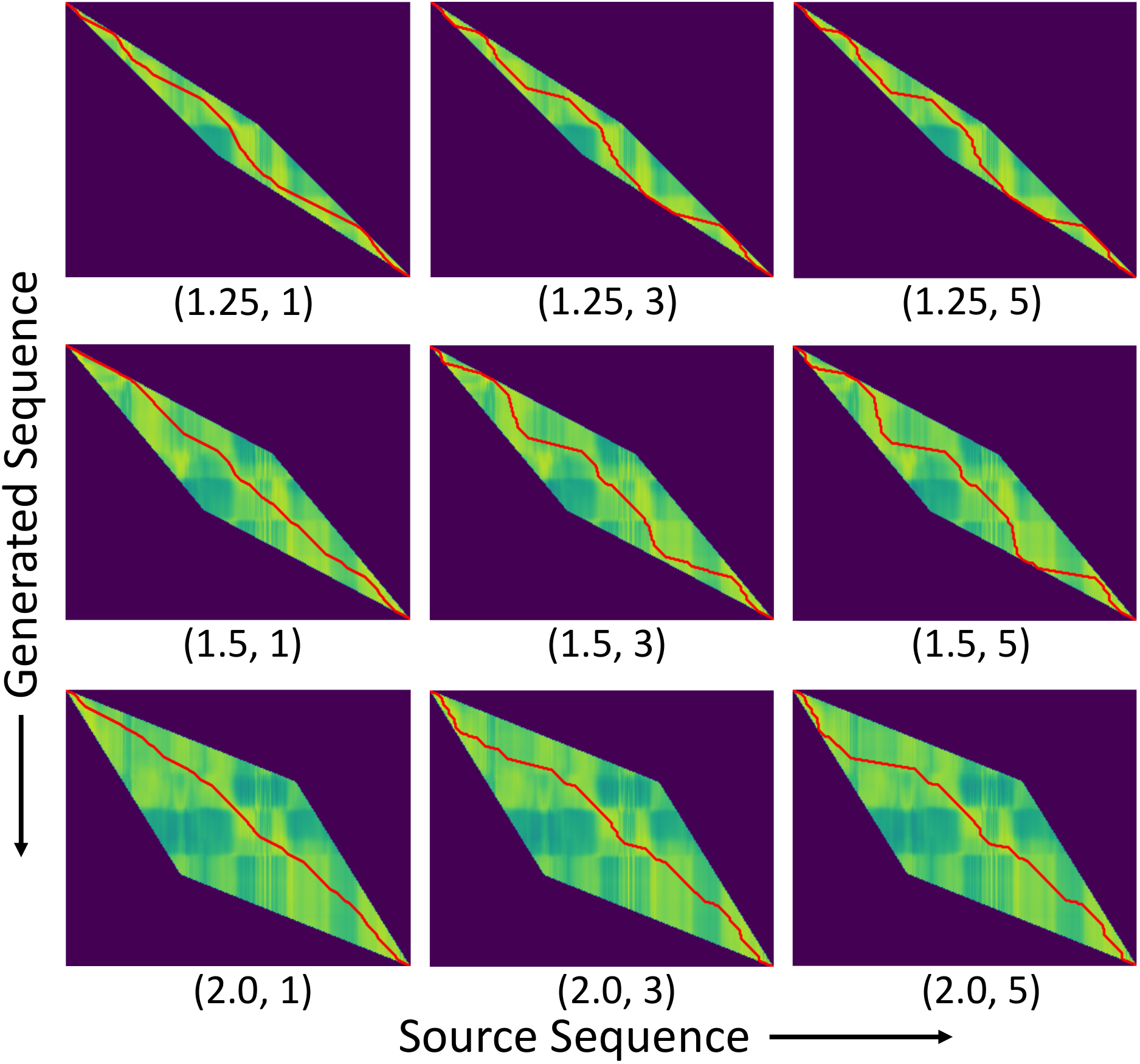}
  \caption{Effect of slope and step constraint on the alignment. The tuple under each image is in (slope, constraint) format. The red curve is the optimal path obtained using back-tracking on the attention maps.}
  \label{fig:slope_steps}
  \vspace{-4mm}
\end{figure}

Fig.~\ref{fig:slope_steps} shows the effect of modifying the slope of the Itakura parallelogram and the horizontal/vertical movement constraint during DTW. As expected, relaxing the slope constraint and increasing the number of horizontal/vertical moves provide more flexibility in adjusting the speaking rate of generated speech. However, this flexibility can lead to missing or distorted phonemes, suggesting a trade-off between changing the speaking rhythm and preserving naturalness. Our framework allows the user to tune these knobs for their own application. 

\vspace{-3.5mm}
\subsection{Reconstruction Quality}
Finally, we crowd source the mean opinion score (MOS) for the re-synthesized speech in the test set using Amazon mechanical turk (AMT). As seen in Fig.~\ref{fig:mos_score}, our framework achieves an average MOS between $3.7-4.0$ across the four tasks. This performance is at par with state-of-the-art neural vocoders trained on hundreds of hours of speech. We note that CMU-ARCTIC task has the lowest MOS, perhaps due to the longer and more complex utterances. Interestingly, the MOS is unaffected by errors in length prediction, as evidenced by the VESUS neutral-angry emotion conversion task. This suggests that our approach of combining the neural network attention weights with a structured DTW algorithm provides robustness to both the speech characteristics and estimation errors. 

\vspace{-2mm}
\section{Conclusions}
We have presented a novel deep-Bayesian framework for adaptive speech duration modification. Our model used a convolutional encoder-decoder architecture to estimate attention maps to associate frames of the input speech with frames of the target. The attention maps are modeled as latent variables, which lead to a stochastic formulation of the MAE loss for model training. During testing, the attention map is directly used to approximate the similarity matrix for a DTW-style backtracking procedure. We evaluated our framework on a voice conversion and three separate emotion conversion tasks. Overall, our framework produces similar duration modification as the vanilla DTW but \textit{without requiring access to the target utterance}. Further, we show that the re-synthesized speech has similar quality to most state-of-the-art neural vocoders.

\bibliographystyle{IEEEtran}

\bibliography{mybib}

\newpage
\section*{Supplemental Materials: \\[1ex]
A Deep-Bayesian Framework for Adaptive Speech Duration Modification}
\vspace{3mm}

\subsection{Loss derivation}
We use a convolutional neural network to predict the length and target speech frame using the following expression:
\begin{equation}
\hat{T} = f_\gamma(X) \;\;\; and \;\;\; \hat{Y}_t = X \cdot A_t + f_\theta(X, \hat{Y}_{0:t-1}).
\label{eqn:T_y_estimation}
\end{equation}
We maximize the log likelihood of the observed data $(X,Y)$ to estimate the weights of the neural network denoted by~$\{\theta,\gamma\}$. This data likelihood can be written as:
\begin{equation}
P(\hat{Y}, \hat{T} | X) = P(\hat{T} | X) \prod_{t=1}^{\hat{T}} P(\hat{Y}_t | X, \hat{T}, \hat{Y}_{0:t-1})
\label{eqn:joint_lkl}
\end{equation}
By expanding the second term of Eq.~(\ref{eqn:joint_lkl}) and using the conditional independece from the graphical model, we have:
\begin{align} \nonumber
P(&\hat{Y}_t | X, \hat{T}, \hat{Y}_{0:t-1}) = \sum_{A_t} P(\hat{Y}_t, A_t | X, \hat{T}, \hat{Y}_{0:t-1}, M) \\ \label{eqn:mask_attention}
& = \sum_{A_t} P(\hat{Y}_t | X, \hat{T}, A_t, \hat{Y}_{0:t-1}) P(A_t | X, \hat{Y}_{0:t-1}, M) 
\end{align}
The attention mask $M$ is a deterministically constructed from the source speech length~$T_s$ and the estimated target length~$\hat{T}$. 

In this work, we encode the attention $A_t$ as a one-hot vector across the $T_s$ input frames of the source speech. Therefore, it follows a multinomial distribution. For simplicity, we model $A_t$ as conditionally independent of the utterance length $T$ given the mask $M$ and the input~$X$. Specifically, taking the $\log(\cdot)$ of Eq.~(\ref{eqn:joint_lkl}) and combining with  Eq.~(\ref{eqn:mask_attention}) yields:

{\footnotesize
\begin{align} \nonumber
\mathcal{L}&(\theta, \gamma) = -\log\big( \sum_{A_t} P(\hat{Y}_t, A_t | X, \hat{T}, \hat{Y}_{0:t-1}, M) \big) - \log\big( P(\hat{T} | X) \big) \\ \nonumber
& = -\log\Big( \sum_{A_t}  \frac{q_\theta(A_t | X, \hat{Y}_{0:t-1}, M)}{q_\theta(A_t | X, \hat{Y}_{0:t-1}, M)} P(\hat{Y}_t, A_t | X, \hat{T}, \hat{Y}_{0:t-1}, M) \Big) \\ \nonumber
& \phantom{\hspace{0.75in}} -\log\big( P(\hat{T} | X) \big) \\ \nonumber
& \leq -\sum_{A_t}q_\theta(A_t|X,\hat{Y}_{0:t-1},M) \log\big( P(\hat{Y}_t | X, A_t, \hat{Y}_{0:t-1}) \big) \\ \nonumber
& \phantom{\hspace{0.75in}} - \log\big( P(\hat{T} | X) + KL(q_\theta(A_t) || P(A_t)) \\ \nonumber
& = -\sum_{A_t}q_\theta(A_t|X,\hat{Y}_{0:t-1},M) \log\big( P(\hat{Y}_t | X, A_t, \hat{Y}_{0:t-1}) \big) \\ \nonumber
& \phantom{\hspace{0.75in}} - \log\big( P(\hat{T} | X) - H(q_\theta) + const. \\ \nonumber
& \leq -\sum_{A_t}q_\theta(A_t|X,\hat{Y}_{0:t-1},M) \log\big( P(\hat{Y}_t | X, A_t, \hat{Y}_{0:t-1}) \big) \\ 
& \phantom{\hspace{0.75in}} - \log\big( P(\hat{T} | X) + const. 
\label{eqn:loss_derivation}
\end{align}}
The distribution~$q_{\theta}(\cdot)$ above is an approximating distribution for the attention vectors implemented by a convolutional network. The first inequality uses the convexity of the $-\log$ function, and the second inequality comes from the fact that entropy $H(q_\theta) \geq 0$. Notice that we have implicitly assumed $P(A_t | X, \hat{Y}_{0:t-1}, M)$ has a uniform distribution over the masked region. This is a reasonable assumption given that the masking process reduces the attention domain to a small region. However, the form of $q_\theta$ is \textbf{not penalized} for deviating from this uniform distribution during training. This flexibility allows the network to learn realistic attention vectors during autoregressive decoding. Eq.~(\ref{eqn:loss_derivation}) can be easily translated into a neural network loss function which we minimize for $\{\theta,\gamma\}$:
\begin{align} \nonumber
L &= E_{A_t \sim q_\theta} \big[ \log \big( P(\hat{Y}_t | X, A_t, \hat{Y}_{0:t-1}) \big) \big] + \log\big( P(\hat{T} | X) \big) \\ \label{eqn:expected_loss}
& = \lambda_1 \times E_{A_t}\big[ \Arrowvert \hat{Y}_t - Y_t^0 \Arrowvert_1 \big] \; + \;  \lambda_2 \times \Arrowvert \hat{T} - T^0 \Arrowvert_1
\end{align}

$\lambda_1$ and $\lambda_2$ are the model hyperparameters that adjusts the trade-off between the two objectives and implicitly contain the variances of the Laplace distributions introduced in the main text. Notice that the loss in Eq.~(\ref{eqn:expected_loss}) computes an expectation over the attention maps. We use the Monte-Carlo estimate by sampling from the attention map at each time-step. The training procedure is therefore stochastic in nature due to this random sampling. We mix this stochastic version with the maximum aposteriori estimate (MAP) of the attention vector with probability of 0.1 in the beginning of training procedure. 

\subsection{Training Algorithm}
\begin{algorithm}
\setstretch{1.1}
    \SetKwInOut{Input}{Input}
    \SetKwInOut{Output}{Output}
    function \underline{trainModelParameters} $(X, Y)$\;
    \Input{filterbank energies ($X \in \mathbb{R}^{D \times T_s}$, $Y \in \mathbb{R}^{D \times T_t}$)}
    \Output{model parameters ($\theta, \gamma$)}
    Set epoch = 0 and threshold\;
    \If{epoch $<$ MaxEpochs}
    {
        Predict length of target sequence $\hat{T} = f_\gamma(X)$\;
        Create attention mask $M \in \mathbb{R}^{T_s \times T_t}$ and set $t = 0$\;
        Estimate $A \in \mathbb{R}^{T_s \times T_t}$ using masked convolution\;
        Sample $u \sim U(0,1)$\;
        \eIf{u $<$ threshold}
        {
            Sample $a \in \mathbb{R}^{T_s \times T_t}$ as 1-hot vectors from $A$\;
            Reconstruct using $\hat{Y}_t = X \cdot a + f_\theta(X, Y_{0:t-1})$\;
        }
        {
            Reconstruct using $\hat{Y}_t = X \cdot A + f_\theta(X, Y_{0:t-1})$\;
        }
        {
        Compute reconstruction and length prediction error\;
        Update parameters $\theta, \gamma$ using backpropagation\;
        epoch $\leftarrow$ epoch + 1\;
        }
    }
    return $\theta$ and $\gamma$\;
    \caption{Strategy for model training}
    \label{alg:training_strategy}
\end{algorithm}
We start with a small threshold in line~8 (i.e., low contribution of the stochastic loss) to prevent the model from diverging in sub-optimal directions. The MAP estimate helps in this regard. Once, the number of training epochs exceed a fix value, we increase threshold to place more emphasis on the stochastic loss. Empirically, we found this to be extremely helpful in generating monotonic attention. We fixed the slope of attention mask in line 5 to $1.25$ based on the relative difference in length observed in the training datasets.

\end{document}